\documentclass[twocolumn,jpa,superscriptaddress,showpacs]{revtex4}
\usepackage{hyperref}
\hypersetup{hypertex=true,
            colorlinks=true,
            linkcolor=red,
            anchorcolor=blue,
            citecolor=blue}
\makeatletter

\newcommand{\Rmnum}[1]{\expandafter\@slowromancap\romannumeral #1@}
\makeatother
\usepackage{graphicx}
\usepackage{subfigure}
\usepackage{bm}
\usepackage{indentfirst}
\usepackage{amsmath} 

\newcommand{\beq}{\begin{eqnarray} }
\newcommand{\eeq}{\end{eqnarray} }
\newcommand{\Beq}{\begin{eqnarray*} }
\newcommand{\Eeq}{\end{eqnarray*} }

\begin{document}

\title{\large \bf Phase diagram for Hole-Doped Kitaev System on the Honeycomb Lattice }  

\author{Su-Ming Zhang}
\affiliation{Renmin University of China, Haidian District, Beijing, China, 100872}

\author{Zheng-Xin Liu}
\thanks{liuzxphys@ruc.edu.cn}
\affiliation{Renmin University of China, Haidian District, Beijing, China, 100872}
\date{\today}

\begin{abstract}

In extended Kitaev models on the honeycomb lattice, off-diagonal interactions (e.g. the $\Gamma, \Gamma^{'}$ terms) can give rise to non-Kitaev quantum spin liquids and several magnetically ordered phases. In the present work, we dope holes to the system and study the resultant $t$-$K$-$\Gamma$-$\Gamma^{'}$ model using mean field theory. The interplay between the charge and spin degrees of freedom results in a rich phase diagram. Similar to doped cuprates, superconductors, pseudogap phases, fermi liquid, strange metal and paramagnetic phase are generated. What is different is that, we obtain more than one superconducting phase (including a topological one) and more than one pseudogap phase no matter what the original spin state is. The Chern number of the topological superconductor is either $\nu=\pm2$ or $\nu=\pm1$, depending on the ratio $\Gamma/ |K|$ in the spin channel. We further find that an intermediate in-plane magnetic field can slightly enlarge the size of the topological superconducting phase. 

\end{abstract}
\maketitle

\section{INTRODUCTION}

Quantum spin liquids (QSLs) are exotic phases of matter which exhibit  long-range entanglements but no traditional symmetry-breaking orders even at extremely low temperatures.\cite{balents2010spin,kitaev2006anyons} The strong quantum fluctuations in systems with geometric frustrations or frustrating interactions causes the melting of long-range magnetic orders, resulting in disordered quantum state without breaking of spin and lattice symmetries. The elementary excitations in a gapped QSL obey fractional statistical and have potential applications in quantum computations\cite{zhou2017quantum,kitaev2006anyons}. Therefore the realization of QSLs has attracted lots of interest in condensed matter physics. Theoretical studies indicate that QSLs generally result from the competition of different magnetic orders\cite{chandra1988possible,he2014chiral,yan2011spin}. Experimentally, candidate QSL materials have been reported on various lattice structures, including triangular lattice, kagome lattice and even three dimensional magnets\cite{shimizu2003spin,okamoto2007spin,han2012fractionalized,chillal2020evidence}. 

QSLs also exist in magnets with strong spin-orbit couplings. In 2006, Kitaev proposed a honeycomb lattice model which has an exactly solvable QSL ground state\cite{kitaev2006anyons}. Later, candidate materials containing the Kitaev type interactions, called the Kitaev materials, are proposed, including the well studied $\alpha$-RuCl$_3$  and Na$_2$IrO$_3$\cite{jackeli2009mott, chaloupka2010kitaev}. However, most of these materials exhibit zigzag type long-range order at low temperatures, indicating the existence of non-Kitaev interactions such as the Heisenberg exchanges\cite{janssen2017magnetization, winter2017breakdown, kimchi2011kitaev, winter2016challenges, singh2012relevance} and off-diagonal $\Gamma,\Gamma'$ terms\cite{winter2016challenges, rau2014trigonal, chaloupka2015hidden, gordon2019theory, lee2020magnetic}. Interestingly, external forces, such as magnetic fields\cite{baek2017evidence,zheng2017gapless,yadav2016kitaev} or mechanical pressure\cite{mirebeau2002pressure,kozlenko2008high,shimizu2003spin,wang2018pressure,cui2017high}, can suppress the order and drive the system into a spin-liquid like or a dimer-like disordered state. 

An intuitional picture for a QSL with continuous spin rotation symmetry is the Resonating Valence Bond(RVB) state raised by Anderson \cite{anderson1987resonating}. In a RVB picture, the spins form singlet pairs and the ground state is a superposition of all possible pairing configurations, resultantly the lattice rotation and translation symmetries are restored. It was also proposed that when doping charge carriers into the system, the previously paired electrons behave like Cooper pairs and move coherently. Therefore, the QSL state is driven into a superconductor \cite{anderson1987resonating}. This provides a scheme to explain the mechanism of high-temperature superconductivity in cuprate oxides\cite{lee1992gauge}. In cuprates, the Neel order is easily destroyed by doped charge carriers. When the doping concentration falls in the optimal region, the cuprate becomes a superconductor with a  high critical temperature\cite{bednorz1986possible, wu1987superconductivity,jiang2018superconductivity,ruan2016relationship,mei2012luttinger,weng2011superconducting}.   

An interesting question is what happens if charge carriers are doped into the Kitaev materials which have anisotropic spin-spin interactions with strong spin-orbit couplings. Anderson's theory indicates that superconductors should appear. This was verified in an earlier study of the $t$-$K$-$J$ model\cite{you2012doping}. In the present work, we dope holes into Kitaev systems with off-diagonal $\Gamma$ and $\Gamma'$ exchange interactions, and study the resultant $t$-$K$-$\Gamma$-$\Gamma'$ model using mean field theory.  The phase diagram is very rich. No matter what the original spin state is,  more than one superconductor and more than one pseudogap phase emerge at different doping concentration. Similar to cuprates,  Fermi liquid phase or the strange metal phase appear at heavy doping. Interestingly, several topological superconductors are obtained, whose Chern number ($\nu=\pm2,\pm1$) are dependent on the ratio $\Gamma/|K|$ in the spin-spin interactions. We further find that intermediate in-plane magnetic fields can enlarge the range of the SC phases. 

The rest part of the present work is organized as follows. In section II, we introduce the effective lattice model (\ref{H2}) as the starting point. In section III, we introduce the slave particle  representation and the related mean field theory  in details, we emphasize the existence of internal gauge symmetry in the slave particle representation and specially introduce the symmetry group of the mean-field description of QSLs --- the projective symmetry groups. In section IV, we present the mean-field phase diagrams at zero doping and finite doping, respectively. Section V is devoted to the conclusions and  discussions.

\section{The Model}

In most  Kitaev materials, the magnetic cations interact with each other via super-exchanges mediated through nonmagnetic anions. Therefore, both the $d$ (or $f$) electrons in the cations and the $p$ electrons in the anions are involved in the electronic structures, thus more than one electronic band enters the low-energy physics. However, we still adopt the single band assumption 
to describe the low energy properties of the system, just like in the cuprate oxides.

To study the effect of charge carriers, we introduce the electron hopping term (namely the $t$ term) in the  $K$-$\Gamma$-$\Gamma'$ effective spin model, in analogy to the $t$-$J$ model for cuprates.

The Heisenberg exchange interactions are believed to be small and are hence neglected in our discussion.  Hence, the starting point of our study is the following effective model,
\begin{align}
H=&-t\sum_{\sigma \langle ij \rangle }(c^\dagger_{i\sigma}c_{j\sigma} + {\rm H.c.})-\mu\sum_ic^\dagger_{i\sigma}c_{i\sigma}\nonumber
\\&+\sum_{\langle ij \rangle \in \alpha \beta (\gamma)} \Big[KS^\gamma_iS^\gamma_j
+\Gamma (S_i^\alpha S_j^\beta + S_i^\beta S_j^\alpha)\nonumber
\\&+\Gamma^{'}(S_i^\alpha S_j^\gamma+S_i^\gamma S_j^\alpha+S_i^\beta S_j^\gamma+S_i^\gamma S_j^\beta)\Big],
\label{H2}
\end{align}
where $t$ is the electron hopping constant, $\mu$ is the chemical potential of the electrons used to control the density of the doped holes, $K$ stands for the Kitaev interaction, $\Gamma$ and $\Gamma'$ are off-diagonal spin-spin interactions(where $\alpha,\beta,\gamma=x,y,z$).  For a singly occupied site $i$, the spin operators $S_i^\alpha$ are related to the electron operators as $S_i^\alpha= C_i^\dag {\sigma_\alpha\over2} C_i$  with $C_i=(c_{i\uparrow}, c_{i\downarrow})$. In later discussions, we assume $K<0, \Gamma>0$ \cite{janssen2017magnetization, liu2018dirac,banerjee2017neutron, wang2017theoretical, cookmeyer2018spin}, and fix $\Gamma'=-0.02|K|$ (the $\Gamma'$ interaction is introduced to stabilize the zigzag order observed in Kitaev materials\cite{winter2016challenges,rau2014trigonal}).

\section{Mean Field Theory}\label{secfer}
\subsection{Slave Particle Representation}

In order to study the above Hamiltonian (\ref{H2}), we introduce the slave particle (called the slave boson) representation to represent the electron creation operators in forms of fermionic spinons $f$ and bosonic holons $b^\dag$, namely
\Beq
c_{i\sigma}=f_{i\sigma}b_i^\dagger,
\Eeq 
where double occupancy is forbidden (otherwise another species of boson called doublon need to be introduced).  The number of the holons $\sum_ib_i^\dag b_i$ equals to the number of doped holes (i.e. the number of removed electrons in the magnetic layer), and the ratio $\delta = {1\over N}\sum_{i} b_i^\dagger b_i$ is the doping concentration relative to half-filling, where $N$ is the total number of sites. 

The spin operators only involve the spinons operators, and can be written as the familiar quadratic form $$S^\alpha_i=\frac{1}{2}f^\dagger_i\sigma_\alpha f_i,$$ where $f^\dagger_i=(f^\dagger_{i\uparrow} , f^\dagger_{i\downarrow})$ and 
$\sigma_\alpha, \alpha=x,y,z$ are the Pauli matrix. For convenience, we introduce a matrix operator 
$\psi_i=  
\begin{pmatrix}
f_{i\uparrow} & f^\dagger_{i\downarrow} \\ f_{i\downarrow} & -f^\dagger_{i\uparrow}
\end{pmatrix}$,
and write the spin operators as $S^\alpha_i={\rm Tr} (\psi^\dagger_i \frac{\sigma_\alpha}{4} \psi_i)$. 

The condition of no double occupancy requires that an on-site particle number constraint should be imposed,
\beq\label{constraint}
\sum_{i\sigma} f_{i\sigma}^\dagger f_{i\sigma} + b_i^\dagger b_i = 1.
\eeq

Obviously, under the phase transformation 
\Beq
f_{i\sigma} \to f_{i\sigma} e^{i\theta_i}, \ \ \ \ b_i^\dag \to b_i^\dag e^{-i\theta_i},
\Eeq
the electron operator $c_{i\sigma}$ remains unchanged $c_{i\sigma} \to c_{i\sigma}$. This reveals an emergent local $U(1)$ gauge symmetry in the slave particle representation of electrons (at half filling the gauge symmetry is enlarged into $SU(2)$ \cite{lee2006doping, affleck19882, wen1996theory, lee19982}). The spinons and the holons interact with each other via the emergent $U(1)$ gauge field. Depending on the electron-electron interactions and the doping concentration, the $U(1)$ gauge field may get confined, resulting in a magnetically ordered phase or a superconducting phase\cite{lee1992gauge}.  To begin with, we need to translate the interactions between the electrons 
into the interactions between the slave particles. 

In the slave-particle representation, the electronic hopping term on a single bond [see (\ref{H2})] can be written in terms of $f_i$ and $b_i$ as
\begin{align}
-t \sum_{\langle ij \rangle \sigma}c_{i\sigma}^\dagger c_{j\sigma} = -t\sum_{\langle ij \rangle \sigma} f_{i\sigma}^\dagger f_{j\sigma} b_j^\dagger b_i .
\label{Ht}
\end{align}
Similarly, the spin-spin interactions can be be written in biquadratic form of spionons. For instance, the Kitaev interaction can be written as 
\Beq
S^m_i S^m_j=-\frac{1}{16}[{\rm Tr} (\psi^\dagger_j \psi_i \psi^\dagger_i \psi_j)
                         +{\rm Tr} (\psi^\dagger_j \sigma^m \psi_i \psi^\dagger_i \sigma^m \psi_j)].
\Eeq 
The fermionic representation of the $\Gamma,\Gamma'$ interactions are given in Appendix \ref{app:dec}. 

An advantage of the slave particle representation is that the various interactions can be treated using mean-field approximation. 

This approach was applied to uncover the mechanism of high-temperature superconductivity in cuprates\cite{lee1992gauge, lee2006doping}. In the rest part of this work, we will study the doped Kitaev systems using mean field theory. 

\subsection{ QSLs at Half-filling }
In mean field theory, the interactions are decoupled into non-interacting Hamiltonians which can be easily solved. Before investigating the effect of hole doping, we firstly introduce the mean field theory of the pure spin-spin interaction,
\beq\label{H1}
H= &&\!\!\!\!\!\!\!\!\!\!\!\!\! \sum_{\langle ij \rangle \in \alpha \beta (\gamma)} KS^\gamma_iS^\gamma_j
+\Gamma (S_i^\alpha S_j^\beta + S_i^\beta S_j^\alpha)\nonumber
\\&\ \ \ &+\Gamma^{'}(S_i^\alpha S_j^\gamma+S_i^\gamma S_j^\alpha+S_i^\beta S_j^\gamma+S_i^\gamma S_j^\beta).
\eeq 
Above Hamiltonian has the $D_{3d} \times Z_2^T$ group symmetry \cite{wang2019one} in the sense that each element is a combination of lattice operation and the corresponding spin operation. Owing  to spin-orbit coupling, the spin axes are locked with the lattice axes. For instance, the $c$-direction of the lattice plane is parallel to ${1\over\sqrt3}(\hat x+\hat y+\hat z)$ direction in the spin-frame, and the three lattice-bond directions are parallel to ${1\over\sqrt2}(\hat x-\hat y)$, ${1\over\sqrt2}(\hat y-\hat z)$ and ${1\over\sqrt2}(\hat z-\hat x)$, respectively.

The model (\ref{H1}) have been studied in literature\cite{wang2020multinode}, and it was known that there are several magnetically ordered phases in the ground state phase diagram. In mean field theory, the magnetically ordered states are essentially the same as the classical ground states whose energy can be easily calculated\cite{rau2014generic}. Therefore, we will focus on possible QSL phases. Noticing that the pure Kitaev model is a special point in the model (\ref{H1}), it is expected that the phase diagram contains at least one QSL phase, the Kitaev Spin Liquid (KSL) phase. In the following we will investigate if there exist other QSL phases.

Since the model (\ref{H1}) only contains nearest neighbor interactions, naturally only nearest neighbor coupling terms need to be considered in the mean field Hamiltonian. Suppose that the QSL contains zero fluxes in each hexagon (it has been shown that the flux-states are not energetically favored\cite{wang2019one}), and that $C_3$ symmetry is preserved in the mean field Hamiltonian.  The most general mean field Hamiltonian takes the following form,
\beq\label{MFtD}
H_{\rm mf}^{\rm SL}\!\!\!&=&\!\!\!\!\!\!\!\!\sum_{\langle ij \rangle \in \alpha \beta(\gamma)}\!\!
\Big\{[f_i^\dagger(t_1^\gamma R_{\alpha \beta} + t_0^\gamma + t_2^\gamma \sigma^\gamma 
+ t_3^\gamma \sigma^\gamma R_{\alpha \beta})f_j \nonumber
\\&&\!\!\! + f_i^\dagger(\Delta_1^\gamma R_{\alpha \beta} + \Delta_0^\gamma 
+\Delta_2^\gamma \sigma^\gamma 
+ \Delta_3^\gamma \sigma^\gamma R_{\alpha \beta})\bar{f}_j] + {\rm H.c.}\Big\} \nonumber
\\&&\!\!\! + \lambda\sum_i  f_i^\dag f_i + {\rm const},
\eeq
where the superscript $\gamma$ labels the orientation of the bond, $R_{\alpha \beta}^\gamma = -\frac{i}{\sqrt{2}}(\sigma^\alpha + \sigma^\beta)$ is a bond-dependent rotation matrix\cite{liu2018dirac}, $\bar{f_i}=(f_{i\downarrow}^\dagger, -f_{i\uparrow}^\dagger)^T.$ $\Delta_{0,1,2,3}^\gamma$ and $t_{0,1,2,3}^\gamma$ are the mean field parameters (complex numbers) representing the fermion pairing and hopping terms, respectively. Owing to the $C_3$ rotation symmetry the values of $t^\gamma$ and $\Delta^\gamma$ are independent on the bond orientation, so we will only consider the $z$-bonds and omit the bond index in later discussion. The self consistent mean-field equations for these parameters read
\beq\label{prmcom}
t_1&=&g_1 {2\over N}\sum_{\langle ij\rangle \in z} \langle f_i^\dagger R_{xy} f_j \rangle^*, \\ 
t_{0,2} &=& g_{0,2}{2\over N}\sum_{\langle ij\rangle \in z} \left[\langle f_i^\dagger \sigma^{x} R_{xy} f_j \rangle^* \pm \langle f_i^\dagger \sigma^{y} R_{xy} f_j \rangle^*\right], \nonumber\\
t_3 &=& g_3{2\over N}\sum_{\langle ij\rangle \in z} \langle f_i^\dagger \sigma^z R_{xy} f_j\rangle^*, \notag\\
\Delta_1 &=& g_1^{'}{2\over N}\sum_{\langle ij\rangle \in z}  \langle f_i^\dagger R_{xy} \bar{f}_j \rangle^*, \nonumber 
\\\Delta_{0,2} &=& g_{0,2}^{'}{2\over N}\sum_{\langle ij\rangle \in z} \left[\langle f_i^\dagger \sigma^{x} R_{xy} \bar{f}_j \rangle^* \pm \langle f_i^\dagger \sigma^{y} R_{xy} f_j \rangle^*\right], \nonumber 
\\ \Delta_3 &=& g_3^{'} {2\over N}\sum_{\langle ij\rangle \in z} \langle f_i^\dagger \sigma^z R_{xy} \bar{f}_j\rangle^*,\notag 
\eeq
where $N$ is the total number of sites, the `coupling constants' $g_{0,1,2,3}$ are related to specific form of the spin-spin interactions and will be specified later. Finally, the Lagrangian multiplier $\lambda$ is introduced to tune the averaged particle number,
\beq\label{N}
1 = {1\over N}\sum_{i} \langle f_i^\dag f_i\rangle.
\eeq
Strictly speaking, the `particle number constraint' contains three components since the SU(2) gauge group have three generators, here we only consider the third one. At half-filling, the self-consistent solution of $\lambda$ is usually zero owing to particle-hole symmetry.

In above mean field theory, we only considered the $C_3$ rotation symmetry and the translation symmetry. Actually, a spin liquid should preserve all the point group symmetries of the spin Hamiltonian. However, in the mean field description of a QSL phase, the symmetry group of the mean-field Hamiltonian is the projective symmetry group (PSG) whose group elements are space group operations followed by $SU(2)$ gauge transformations\cite{wen2002quantum}. We will restrict to the PSG of the exactly solvable Kitaev point (called the Kitaev PSG) \cite{you2012doping, wang2019one} in the following discussion.

To analyze the PSG symmetry, it is helpful to rewrite the mean-field Hamiltonian (\ref{MFtD}) in a matrix form,
\beq\label{MFU}
H^{\rm SL}_{\rm mf}\!\!\!&=&\!\!\!\!\!\sum_{\langle i,j \rangle \in \alpha \beta (\gamma)}\!\!\! {\rm Tr}[U^{(0)}_{ji} \psi^\dagger_i \psi_j]
+{\rm Tr} [U^{(1)}_{ji} \psi^\dagger_i(iR^\gamma_{\alpha \beta})\psi_j] \notag
\\ &&+ {\rm Tr} [U^{(2)}_{ji} \psi^\dagger_i \sigma^\gamma \psi_j]
+{\rm Tr} [U^{(3)}_{ji} \psi^\dagger_i \sigma^\gamma R^\gamma_{\alpha \beta} \psi_j] + {\rm H.c.} \notag
\\ && + \lambda \sum_i {\rm Tr} (\psi_i{\tau^z\over4}\psi_i^\dag ) + {\rm const},
\eeq
where $U_{ji}^{(0,1,2,3)}$ are the matrix form of the mean-field parameters $\Delta_{0,1,2,3}, t_{0,1,2,3}$. To satisfy the Kitaev PSG, only eight real mean field parameters remain, namely, $\eta_{0,1,2}$, $\rho_{a,b,c,d}$ and the Lagrangian multiplier $\lambda$. These eight parameters are real because they are expected values of quadratic majorana fermion operators, for details see Appendix \ref{app:psg}. The matrices $U^{(0,1,2,3)}_{ji}$ can be expanded by these parameters as
\begin{align}\label{prmrel}
&U_{ji}^{(0)}=i\eta_0 + i(\rho_a+\rho_b), \nonumber
\\&U_{ji}^{(1)}=i\eta_1(\tau^x\! +\! \tau^y \!+\! \tau^z)
+i(\rho_a \!-\! \rho_b\! +\! \rho_c \!+\! 2\rho_d)(\tau^\alpha \!+\! \tau^\beta), \nonumber
\\&U_{ji}^{(2)}=i(\rho_a\! +\! \rho_b) \tau^\gamma + i\rho_d(\tau^\alpha \!+\! \tau^\beta)
+i\eta_2(\tau^x \!+\! \tau^y \!+\! \tau^z), \nonumber
\\&U_{ji}^{(3)}=i(\rho_b \!-\! \rho_a \!+\! \rho_c)(\tau^\alpha - \tau^\beta),
\end{align}
where $\tau^{x,y,z}$ are Pauli matrices generating the $SU(2)$ gauge group. 

Comparing (\ref{MFtD}) and (\ref{prmcom}) with (\ref{MFU}) and (\ref{prmrel}), we can express $\eta_{0,1,2}$ and $\rho_{a,b,c,d}$ by the parameters $t_{0,1,2,3}, \Delta_{0,1,2,3}$, namely,
\Beq
 &&\eta_0=-i(t_0-\Delta_2)-it_2(1-i), \nonumber
 \\&&\eta_1=i(t_1-\Delta_1), \nonumber
 \\&&\eta_2=-it_2, 
 \Eeq
 and  
 \Beq
 \\&&\rho_a=\frac{i}{2}(1-i)t_2+\frac{i}{2}(-t_1-t_3+\Delta_1-\Delta_2),
 \\&&\rho_b=\frac{i}{2}(1-i)t_2+\frac{i}{2}(t_1+t_3-\Delta_1-\Delta_2), \nonumber
 \\&&\rho_c=i(-t_1+t_3+\Delta_1), \nonumber
 \\&&\rho_d=\frac{i}{\sqrt 2}\{[ e^{-i{\pi\over4}}t_0+e^{i{\pi\over4}}t_2+e^{i{\pi\over4}}\Delta_1-e^{-i{\pi\over4}}\Delta_3]-{\rm h.c.}\}.\notag
\Eeq

 Substituting (\ref{prmcom}) and (\ref{N}) into above relations, we obtain the self-consistent mean-field equations for the set of eight real parameters $\eta_{0,1,2},\rho_{a,b,c,d}$, $\lambda$, where the coupling constants $g_{0,1,2,3}$ are given by
\Beq
&&g_0=g'_0=-\frac{1}{8}\Gamma, \\
&&g_1=-\frac{1}{8}(2\Gamma+|K|+4\Gamma^{'}), \qquad g_1^{'}=-\frac{1}{8}(2\Gamma+|K|),\\
&&g_2=-\frac{1}{8}\Gamma, \qquad g_2^{'}=-\frac{1}{8}(\Gamma+4\Gamma^{'}), \\
&&g_{3}=g'_{3}=-\frac{1}{8}|K|.
\Eeq

Solving the above equations, we obtain a family of QSL solutions, two of which appear in the phase diagram (see Fig.\ref{fig:spin} in Sec. \ref{sec:phsdiag}).

\subsection{Charge-doped QSLs}\label{sec:dopQSL}

In this section, we consider the situations in which holes are doped into the QSLs. When holes are present, the electrons can lower their energy by hoping to the vacant sites. From the slave particle representation (\ref{Ht}), we can perform the following mean field decoupling to the electron hoping term,
\beq\label{Htm}
H_{\rm mf}^{\rm hop}\! =\! -t \sum_{\langle ij \rangle, \sigma}\! \Big(\langle f_{i\sigma}^\dagger f_{j\sigma} \rangle b_j^\dagger b_i
+ \langle b_j^\dagger b_i \rangle f_{i\sigma}^\dagger f_{j\sigma} + {\rm H.c.} \Big).
\eeq

In above expression, the boson operators $b, b^\dag$ enters the mean field theory. We need to introduces one more parameter, namely, 
\Beq
\chi_b^\gamma=-t{2\over N}\sum_{\langle ij\rangle \in \gamma}\langle b_i^\dagger b_j \rangle. 
\Eeq
Another parameter $\chi^\gamma=-t{2\over N}\sum_{\langle ij\rangle \in \gamma,\sigma}\langle f_{i\sigma}^\dagger f_{j\sigma}\rangle $ is not new since it is proportional to $t_0$ with $\chi^\gamma = -{t\over g_0}t_0$. Owing to the $C_3$ rotation symmetry, the parameters take the same values on different bonds $\chi^x=\chi^y=\chi^z$ and $\chi^x_b=\chi^y_b=\chi^z_b$, so the superscript $\gamma$ will be omitted later.

Combining of spinon part and the chargon part, we obtain the total mean field Hamiltonian,
\[
H_{\rm mf}^{\rm tot} = H_{\rm mf}^{\rm hop} + H_{\rm mf}^{\rm SL} + \lambda \sum_i  (f_i^\dagger f_i+b_i^\dagger b_i) - \mu \sum_i b_i^\dagger b_i.
\] 
Since we are more interested in the region where the magnetic orders are destroyed by the charge fluctuations(see \ref{sec:Dord} for discussion), we do not consider the magnetic order in the mean field Hamiltonian.  Many ansatz have been tried in determining the phase diagram.

At light doping, namely when $\delta$ is small, the spin-spin interactions play important roles in the low energy physics. 
The ansatz (\ref{MFU}) preserving the Kitaev PSG is lowest in free energy at small $\delta$. With the increasing of $\delta$, the charge part is more important, we find that the system favors a general QSL ansatz (\ref{MFtD}) without the restriction of the Kitaev PSG. Especially, when $\delta$ is very large, the system becomes metalic and the free energy favors the ansatz in which the paring parameters $\Delta_{0,1,2,3}$ vanish and the hoping parameters $t_{0,1,2,3}$ are nonzero.

As an example, we write down the total mean field Hamiltonian in the light doping region. From (\ref{MFU}) and (\ref{Htm}), we obtain, 
\beq\label{Htol}
H_{\rm mf}^{\rm tot}&&= \sum_{\langle ij \rangle \in \alpha \beta (\gamma)}
\big( \chi^*b_i^\dagger b_j  + \chi_b^* f_i^\dagger f_j + {\rm H.c.} \big) \nonumber
\\&&+\sum_{\langle i,j \rangle \in \alpha \beta (\gamma)} 
\Big\{{\rm Tr} [U^{(0)}_{ji} \psi^\dagger_i \psi_j]
+{\rm Tr} [U^{(1)}_{ji} \psi^\dagger_i(iR^\gamma_{\alpha \beta})\psi_j] \nonumber
\\&& \qquad \quad + {\rm Tr} [U^{(2)}_{ji} \psi^\dagger_i \sigma^\gamma \psi_j]
+{\rm Tr} [U^{(3)}_{ji} \psi^\dagger_i \sigma^\gamma R^\gamma_{\alpha \beta} \psi_j] \Big\} \nonumber
\\&& + \quad \lambda \sum_i  (f_i^\dagger f_i+b_i^\dagger b_i) - \mu \sum_i b_i^\dagger b_i + {\rm const}.
\eeq
Notice that the mean field parameter $\chi_b = -t\langle b_i^\dagger b_j \rangle$ (average of boson kinetic term) is generally a complex number which spontaneously breaks the time reversal symmetry. $\lambda$ is a Lagrange multiplier for the particle number constraint (\ref{constraint}), and $\mu$ is the chemical potential to adjust the doping concentration $\delta = {1\over N}\sum_{i} b_i^\dagger b_i$.

At half-filling, namely as $\delta=0$, Eq. (\ref{Htol}) reduces to the mean field description of QSLs with the condensation of the spinon pairs. Each spinon  pair carries 2 units of gauge charge, thus the condensation of the spinon pairs will gap out the $U(1)$ gauge field via Anderson-Higgs mechanism. Resultantly, in the low-energy limit the gauge fields are described by $Z_2$ gauge theory.
Upon finite doping, once the bosonic holons undergoes Bose-Einstein condensation at low temperatures, the $Z_2$ gauge field is further confined. The confinement of the $Z_2$ gauge charge indicating that unpaired spinons cannot be separated, only paired spinons are allowed low-energy excitations.  On the other hand, the $U(1)$ gauge `symmetry' can never be broken, only the $U(1)$ gauge invariant quantities can have nonzero expectation values \cite{ElitzurPRD1975}. 
To obtain a $U(1)$ gauge invariant quantity, one can combine a pair of spinons (which are confined with each other) and two holons, which are essentially equivalent to two electrons, to form a Cooper pair. The $U(1)$ gauge invariance of the Cooper pair can be seen recalling that each spinon $f_i$ carries $+1$ gauge charge and each holon $b_i^\dag$ carries $-1$ gauge charge such that a Cooper pair carries zero gauge charge. The coherent moving of Cooper pairs gives rise to a superconductor. The nonzero expectation value of the Cooper pair in the ground state can be considered as the order parameter of the superconducting phase.

Different from the $d$-wave superconductors in cuprates, here the spinon pairing channel includes both singlet pairing and triplet pairing. Therefore, the pairing symmetry is neither $s(d)$-wave or $p(f)$-wave, and the spatial inversion symmetry is spontaneously broken.

\subsection{Magnetically ordered phases} \label{sec:Dord}

{\it At half-filling.} At zero temperature and at half-filling, the mean field description of the magnetic ordered states is essentially the classical ground state since the magnetic order are generally fully saturated.  For this reason, 
to estimate the ground state energy in the ordered phase, 
we just use the classical method (such as single-Q approximation\cite{rau2014generic}) instead of taking mean field theory. 

To estimation the critical temperature, we adopt the mean-field Hamiltonian (\ref{MFU}) with a background field $\pmb M_i$ which induces the magnetic order. Namely, we add an extra term $H_{mf}=\frac{1}{2} \sum_i \bm{M}_i \cdot f_i^\dagger \bm{\sigma} f_i$ into  (\ref{MFU}). The ordering pattern $\bm{M}_i$ is determined by the classical solution within the single-$Q$ approximation, and the magnitude $M=|\pmb M_i|$ is treated as a variational parameter determined by minimizing the free energy. The threshold temperature at which $M$ reduces to zero is the critical temperature $T_{c\rm M}$. 
In this approach, the spin-spin interactions are decoupled in two different channels, namely  the spin-liquid channel and the magnetically ordered channel, both being included  in the mean field Hamiltonian. Therefore, when calculating the free energy of the system from mean-field theory, we have counted the contribution from the spin-spin interactions twice. This may result in an overcounting of the free energy and may introduce a systematic error in estimating the critical temperature $T_{c\rm M}$.


{\it Finite doping at zero temperature $T=0$}. In the absence of doping, the energy per site in the classically ordered state is equal to half of the total energy of the three bonds connected to it. When the doping concentration is nonzero $\delta\neq 0$, we assume that the holes are uniformly distributed among the lattice sites (namely, we assume that the holes are well separated). Then we approximately count the total energy of the ordered phase by assuming that the each hole increases the total energy by the amount which the interactions on three neighboring bonds would have if there were no holes. 

On the other hand, at finite doping the superconducting states are competing in energy.  When the energy of the ordered state meets that of a superconducting state, then the critical concentration $\delta_{c\rm M}$ is obtained.

\begin{figure}[t]
\setcounter{subfigure}{0}
\centering
\setcounter{subfigure}{0}
\subfigure[phase diagram]{\includegraphics[width=7.cm]{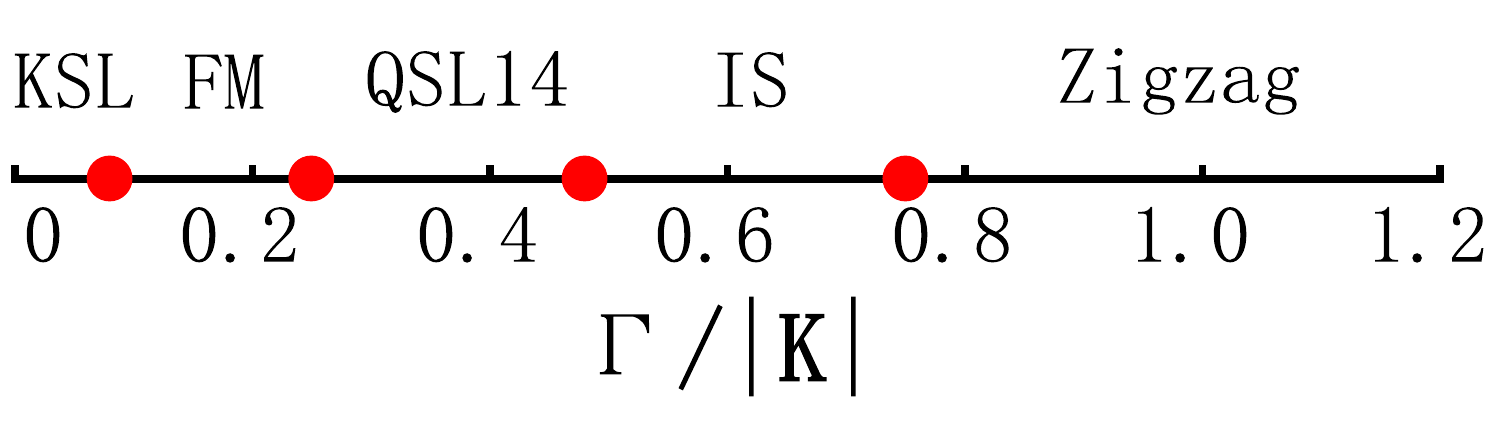}}\\
\subfigure[FM]{\includegraphics[width=2.3cm]{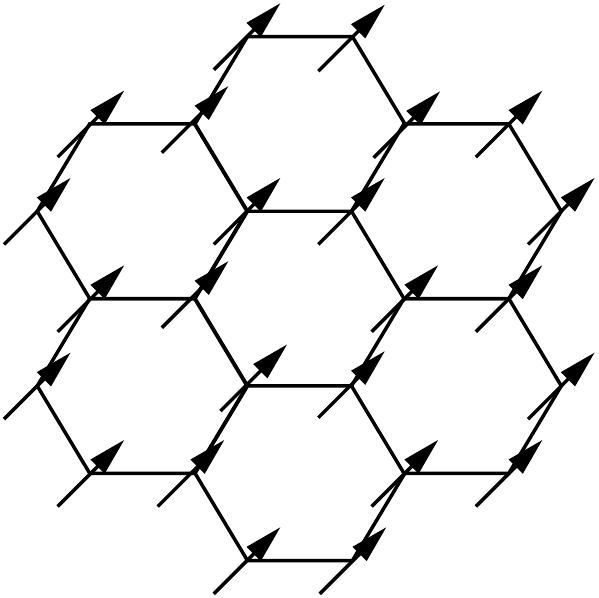}}\ 
\subfigure[Zigzag]{\includegraphics[width=2.3cm]{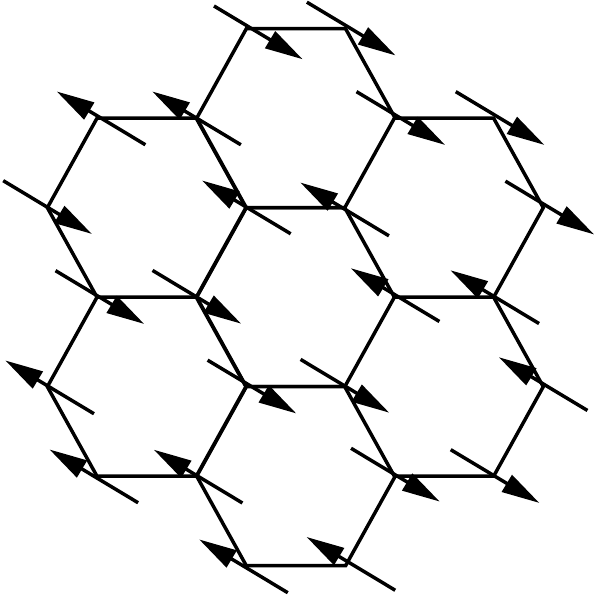}}\ 
\subfigure[spinon dispersion]{\includegraphics[width=3.2cm, height=2.4cm]{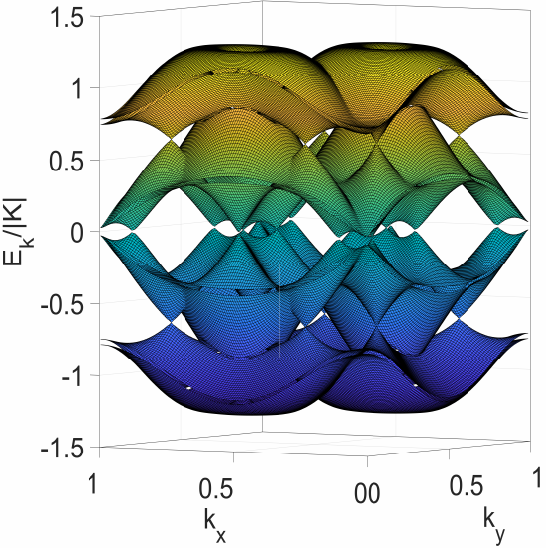}}
\caption{(a) Phase diagram at half-filling with fixed $\Gamma^{'}/|K|=-0.02$.  KSL denotes the Kitaev spin liquid, FM the ferromagnetic phase,QSL14 the 14-cone QSL phase, IS the incommensurate spiral phase, and Zigzag the zigzag ordered phase. (b) and (c) show the spin configurations of the FM and the Zigzag order, respectively. (d) The spinon dispersion in the QSL14 phase.}
\label{fig:spin}
\end{figure}

\section{The PHASE DIAGRAMs} \label{sec:phsdiag}

The phase diagram is obtained by minimizing the free energy. We have tried different self-consistent solutions of the mean field equations. For a given set of interaction parameters, the one with the lowest free energy (or energy at zero temperature)  defines the phase in the phase diagram.

\subsection{Quantum Phase Diagram ($T=0$) at Half-filling}

The spin model with $K$-$J$-$\Gamma$-$\Gamma'$ interactions have bee profoundly studied. Since most phases are magnetically ordered, we will focus on possible QSLs. In our mean field theory,  we indeed obtain two QSL phases. Besides the well known KSL, we obtain an additional gapless QSL phase with 14 majorana cones in the spinon excitation spectrum(a similar 14-cone QSL state called PKSL14 was obtained in Ref.\cite{wang2019one}). The zero-temperature mean field phase diagram is shown in Fig.\ref{fig:spin}.

At the small $\Gamma$ side, the KSL locates in the region $0\leq \Gamma/|K|\leq 0.08$. The KSL with $K<0$ has a ferromagnetic instability. At larger $\Gamma$ with  $0.08<\Gamma/|K|<0.25$, a ferromagnetic(FM) phase is obtained \cite{wang2019one,wang2020multinode,lee2020magnetic}.

At the large $\Gamma$ side, the system favors the zigzag order for $\Gamma/|K|>0.78$. The zigzag order has been observed in serval Kitaev candidates\cite{sears2015magnetic,johnson2015monoclinic,cao2016low,ye2012direct,choi2012spin,williams2016incommensurate}. It was shown that the zigzag phase is absent in the $K$-$\Gamma$ model, for this reason, we introduce a small $\Gamma'=-0.02|K|$ to stabilize it. With the decreasing $\Gamma$, the zigzag order is replaced by an incommensurate spiral (IS) phase at $0.48<\Gamma/|K|<0.78$. 

Finally, resulting from the competition between the FM phase and the IS phase, there exist another interesting QSL phase (labeled as QSL14) living in the region $0.25<\Gamma/|K|<0.48$. This QSL contains 14 majorana cones (see Fig.\ref{fig:spin} (d)) and shares the same PSG as the KSL phase. Although a similar state was obtained from variational Monte Carlo method, the two 14-cone states are different. When a small magnetic field is applied along $\hat c={1\over \sqrt3}(\hat x+\hat y+\hat z)$ direction, the 14-cone QSL from self-consistent solution is turned into a gapped state with Chern number $\nu=-1$, but the PKSL14 acquires a Chern number $\nu=5$ under the same magnetic field. In this region, there is another spin liquid solution  with eight cones (labeled as QSL8), but with a higher energy.

\begin{figure*}[t]
\setcounter{subfigure}{0}
\subfigure[]{\includegraphics[width=4.3cm]{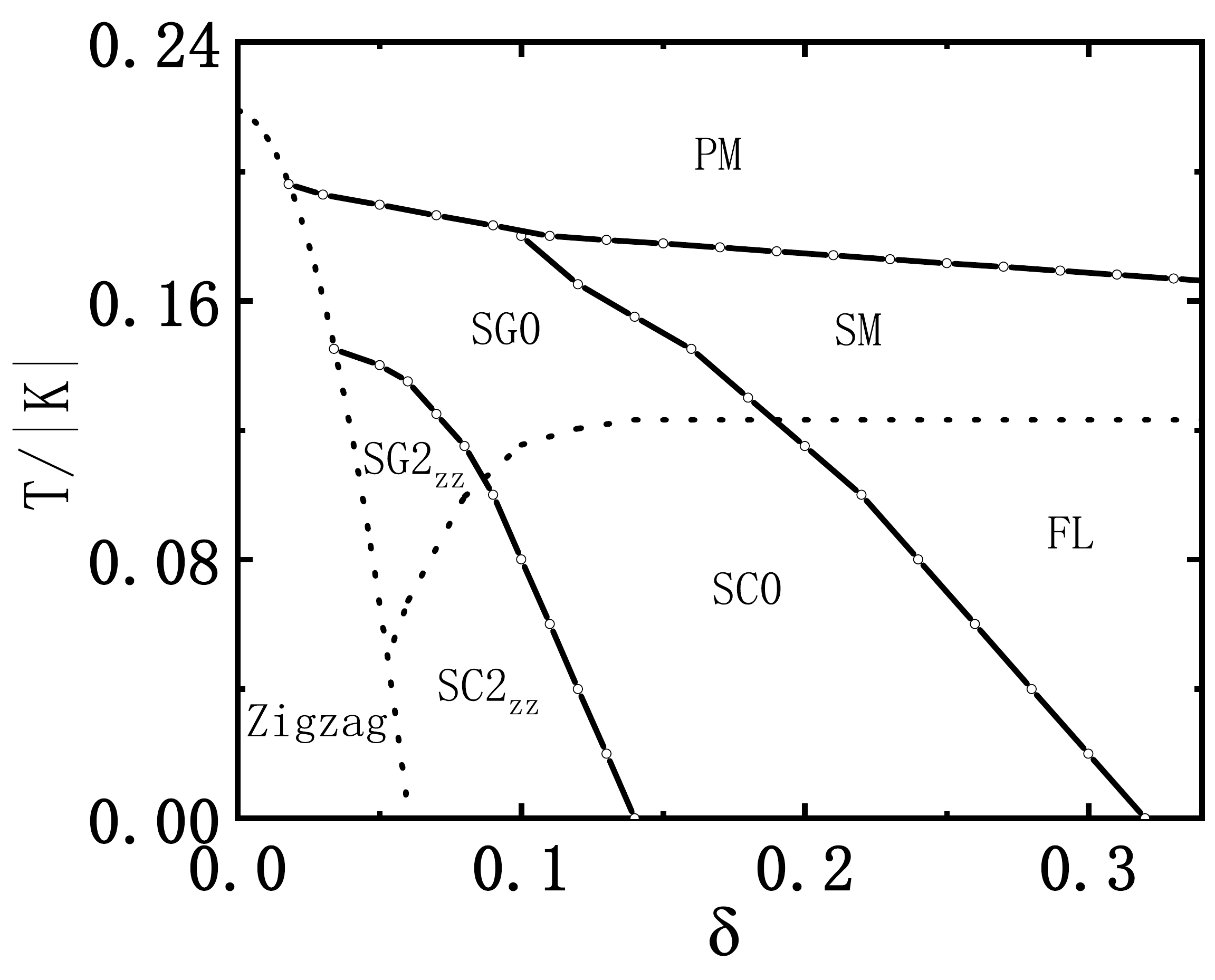}}
\subfigure[]{\includegraphics[width=4.3cm]{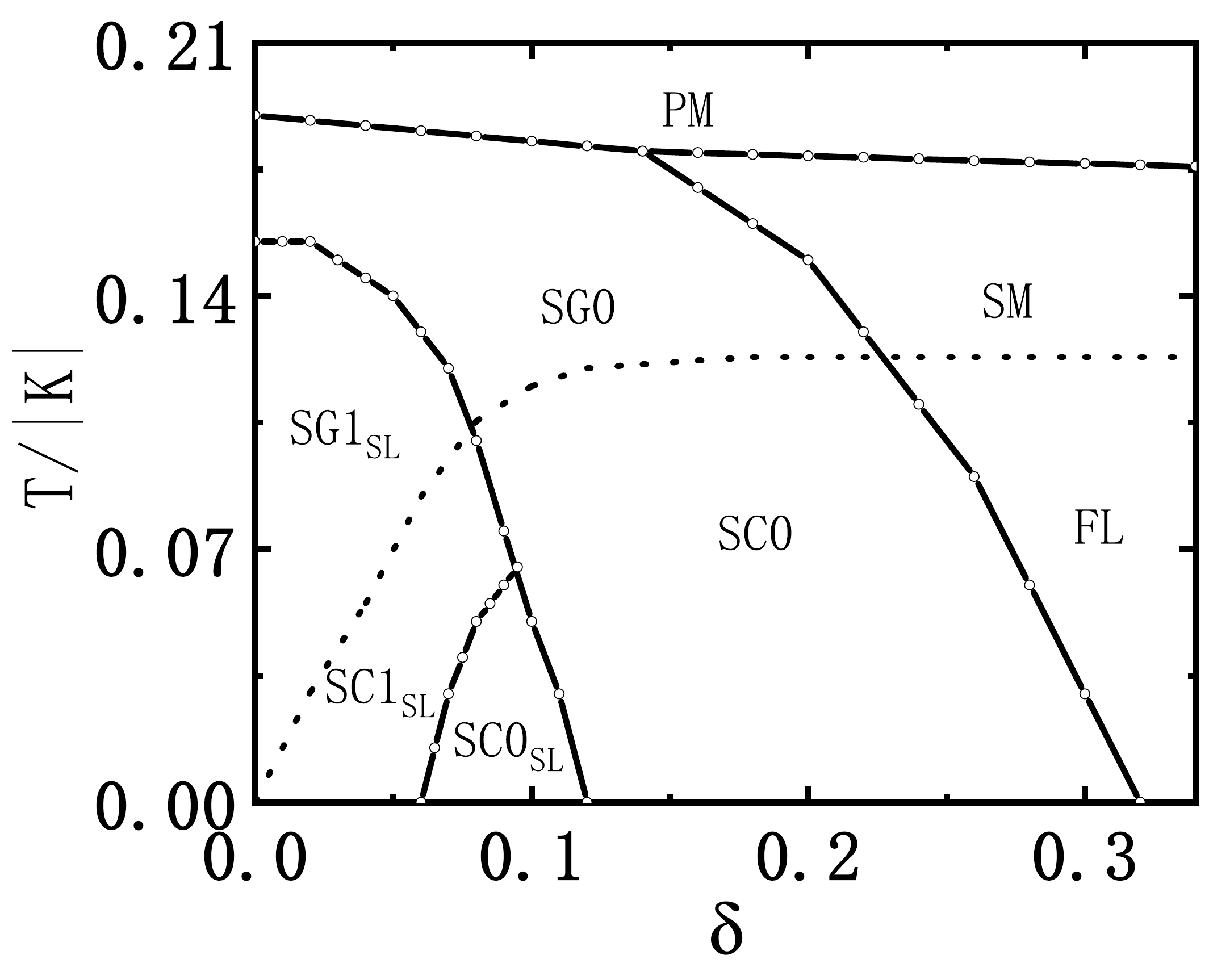}}
\subfigure[]{\includegraphics[width=4.4cm]{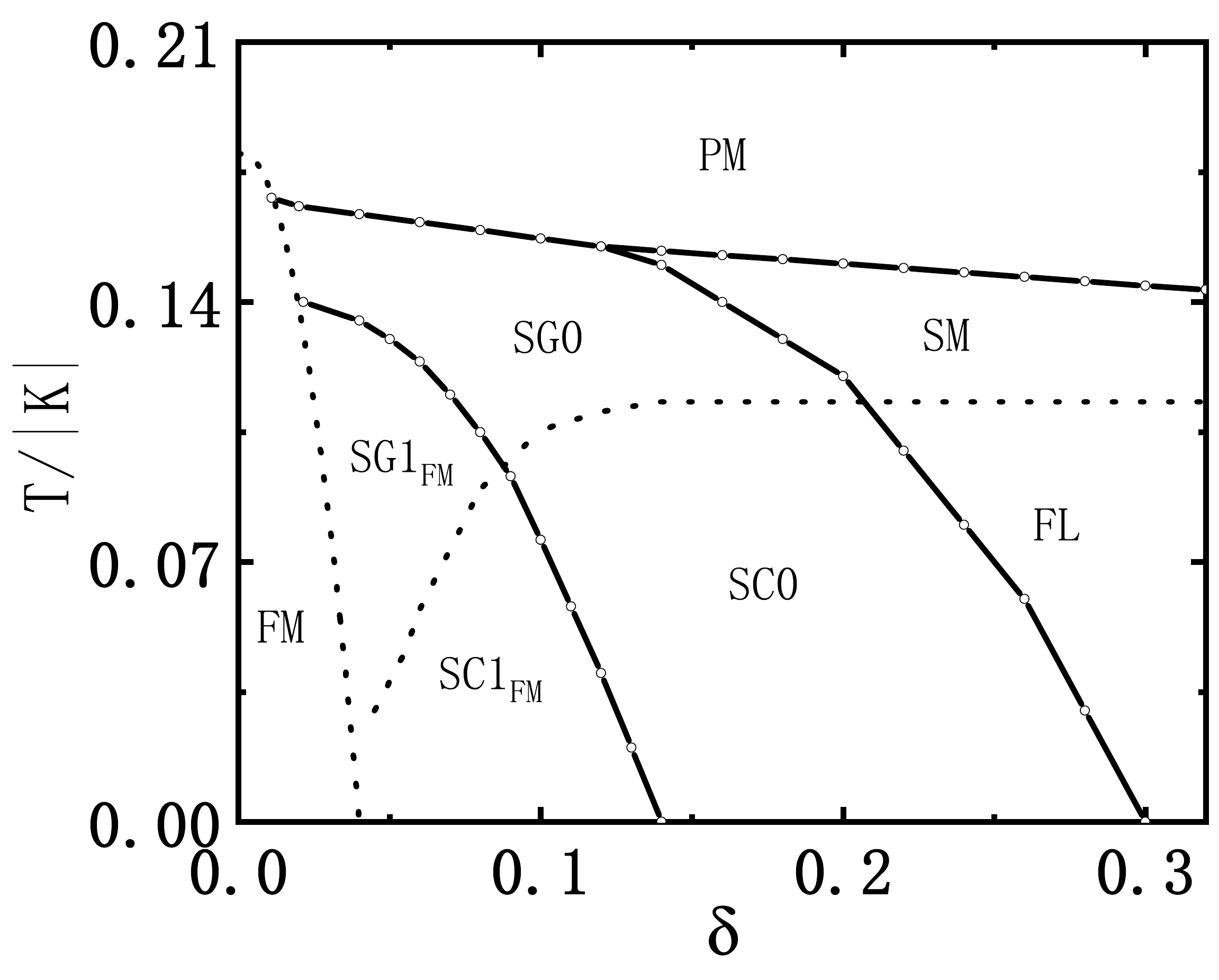}}
\subfigure[]{\includegraphics[width=4.3cm]{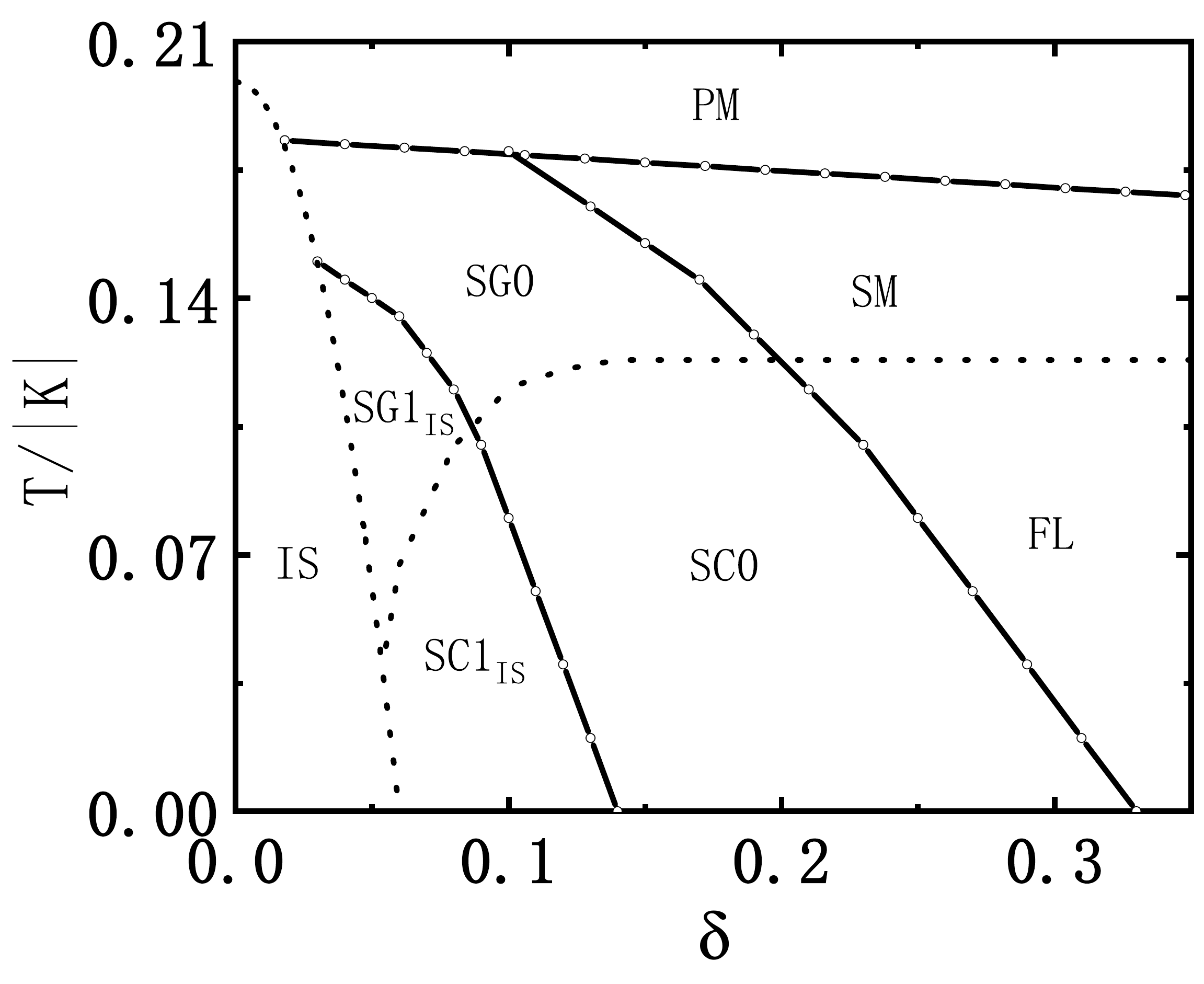}}
\caption{Phase diagrams of doped Kitaev system (with fixed $\Gamma'/|K|=-0.02$). (a) doped zigzag phase with $\Gamma/|K|=1$; (b) doped QSL14 phase with $\Gamma/|K|=0.4$; (c) doped FM phase with $\Gamma/|K|=0.15$; (d) doped IS phase with $\Gamma/|K|=0.6$. SC denotes the superconducting phase, SC0 means the Chern number is $\nu=0$, and SC2$_{\rm zz}$ stands for a SC phase with Chern number $\nu=2$ doped from the zigzag phase, so on and so forth. SG denotes the spin gapped phase, FL the fermi liquid phase, SM the strange metal phase and PM the paramagnetic phase. 
}
\label{figure2}
\end{figure*}

\begin{figure}[b]
\setcounter{subfigure}{0}
\centering
\subfigure[]{\includegraphics[height=3.2cm]{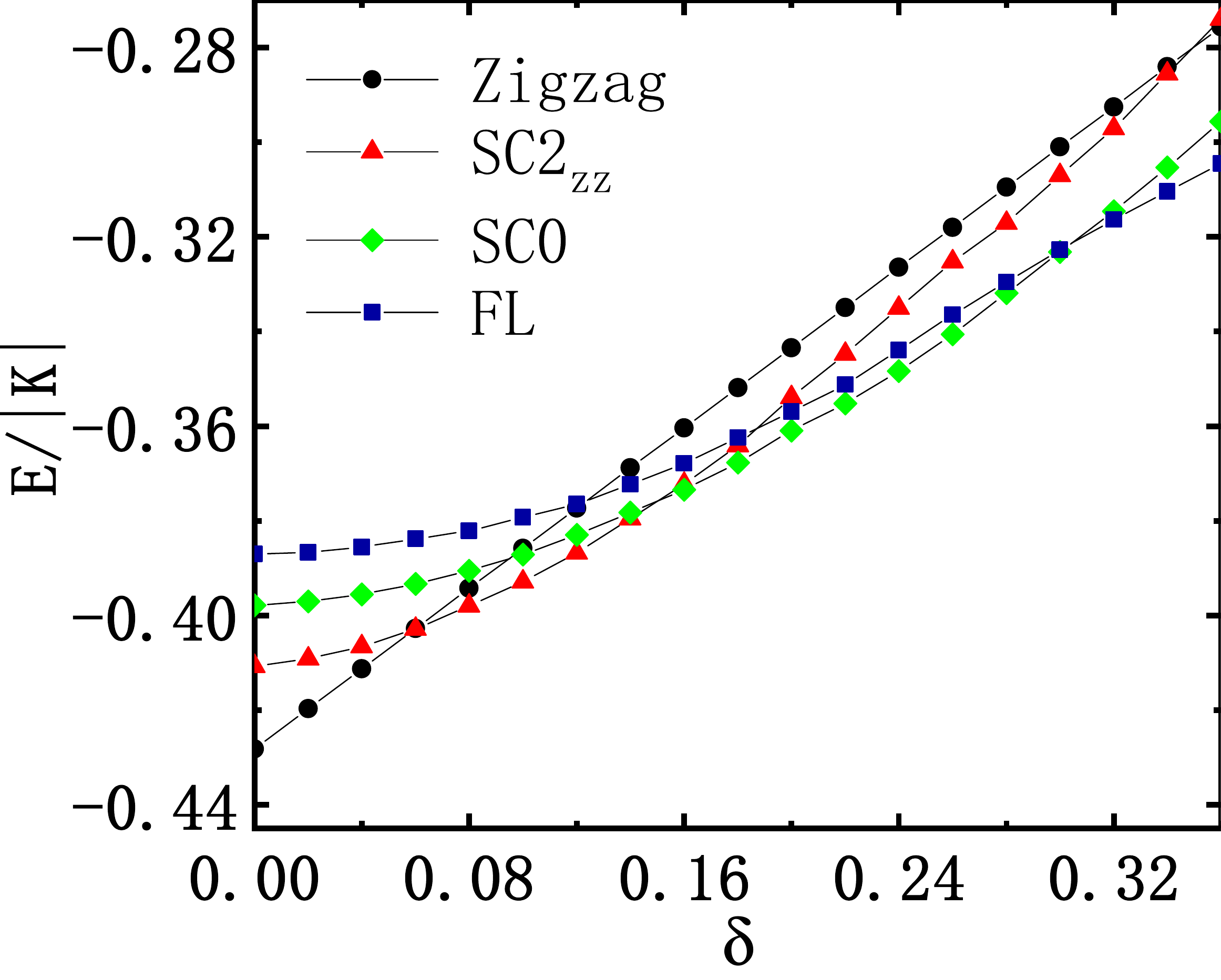}}
\subfigure[]{\includegraphics[height=3.2cm]{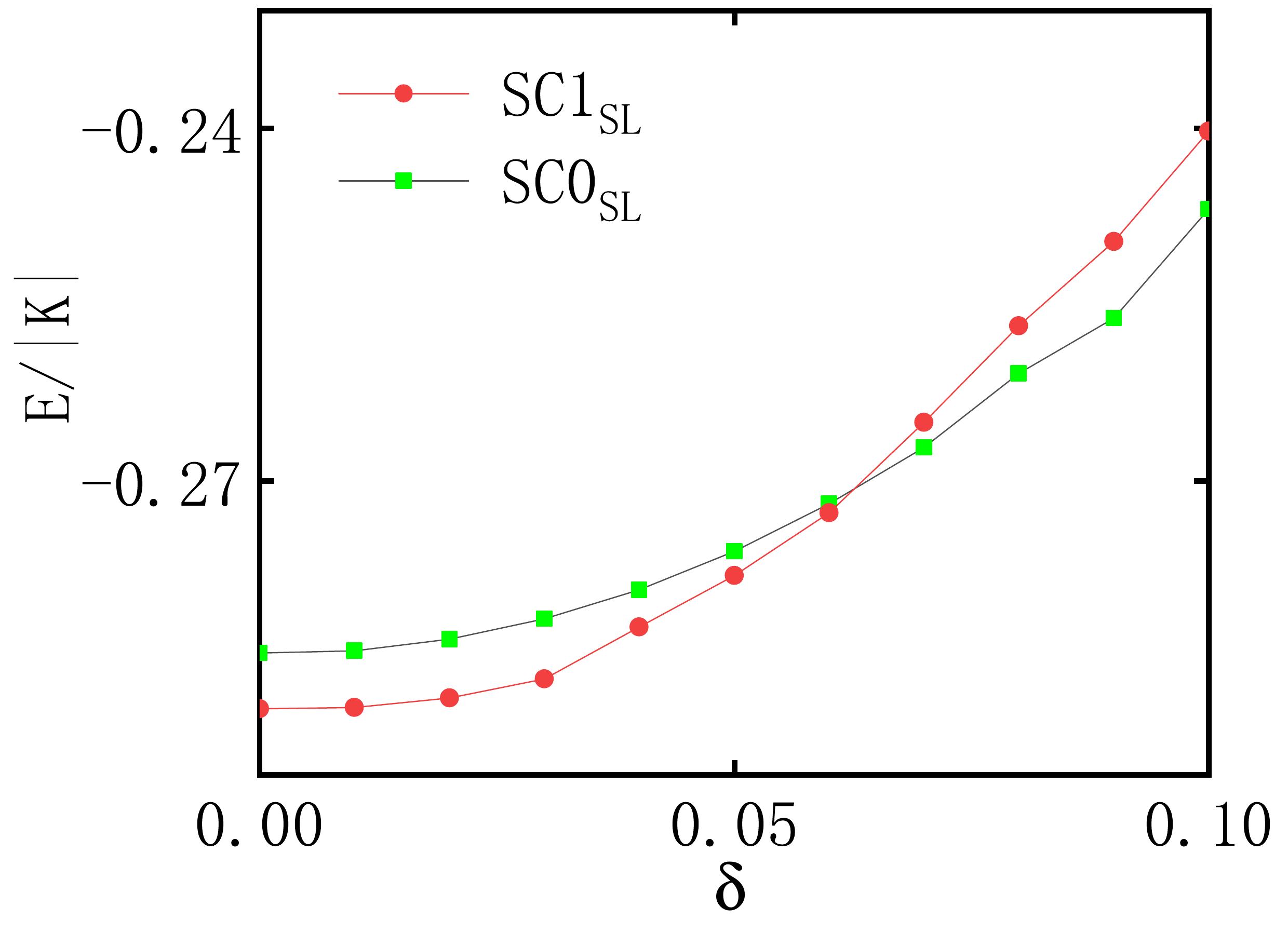}}
\caption{First order phase transitions with increasing $\delta$ at $T=0$ illustrated by the Energy curves. (a) Doped zigzag phase with $\Gamma/|K|=1, \Gamma^{'}/|K|=-0.02$. The level crossings indicate a series of phase transitions from the Zigzag phase to the SC2$_{zz}$, from the SC2$_{zz}$ to the SC0 and from the SC0 phase to the FL. (b) Doped QSL14 with $\Gamma/|K|=0.4, \Gamma^{'}/|K|=-0.02$. The level crossing indicates the transition from the SC1$_{\rm SL}$ phase to the SC0$_{\rm SL}$ phase.  }
\label{fig:trans}
\end{figure}

\begin{figure}[b]
\subfigure[]{\includegraphics[width=4.3cm]{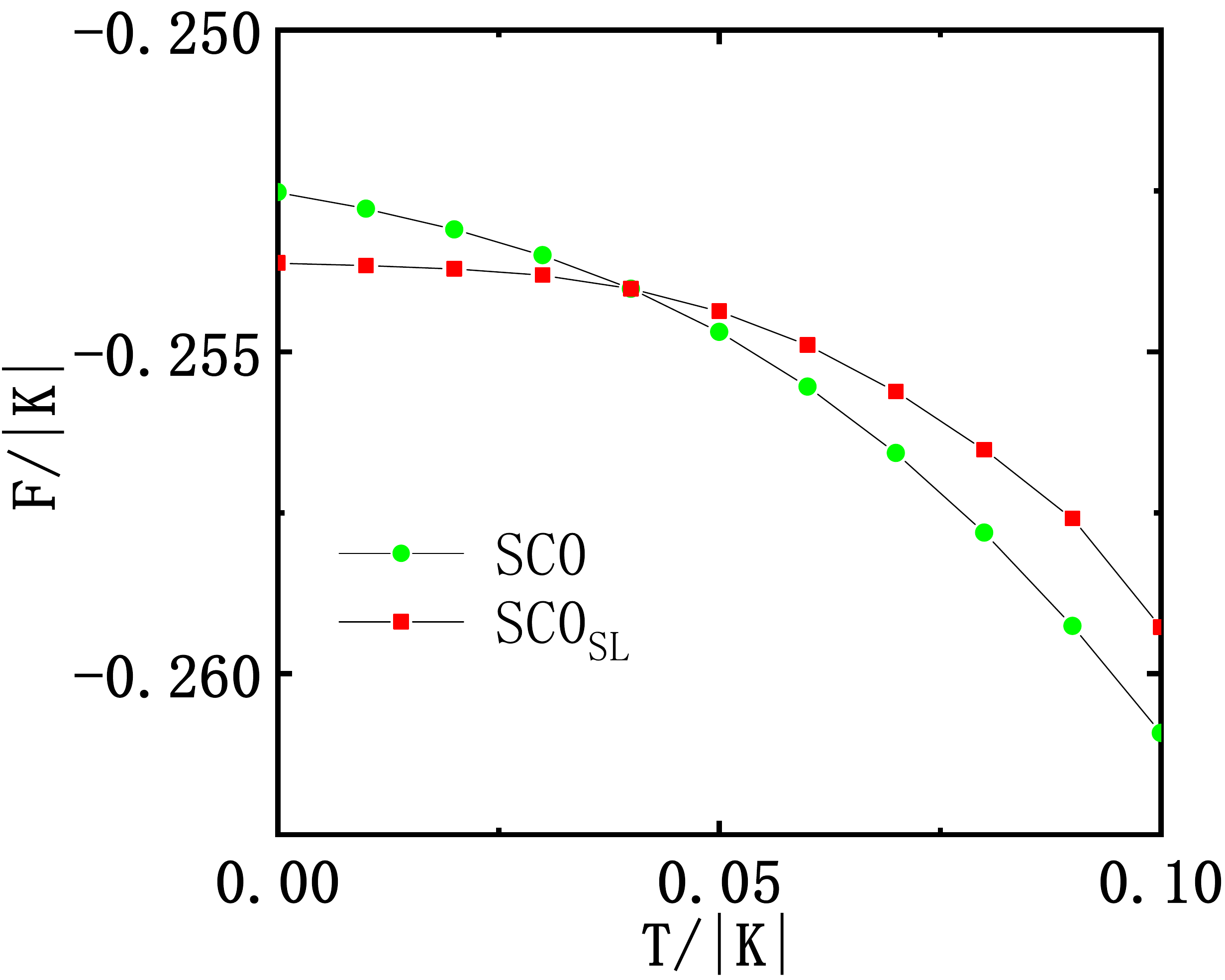}}
\subfigure[]{\includegraphics[width=4.2cm]{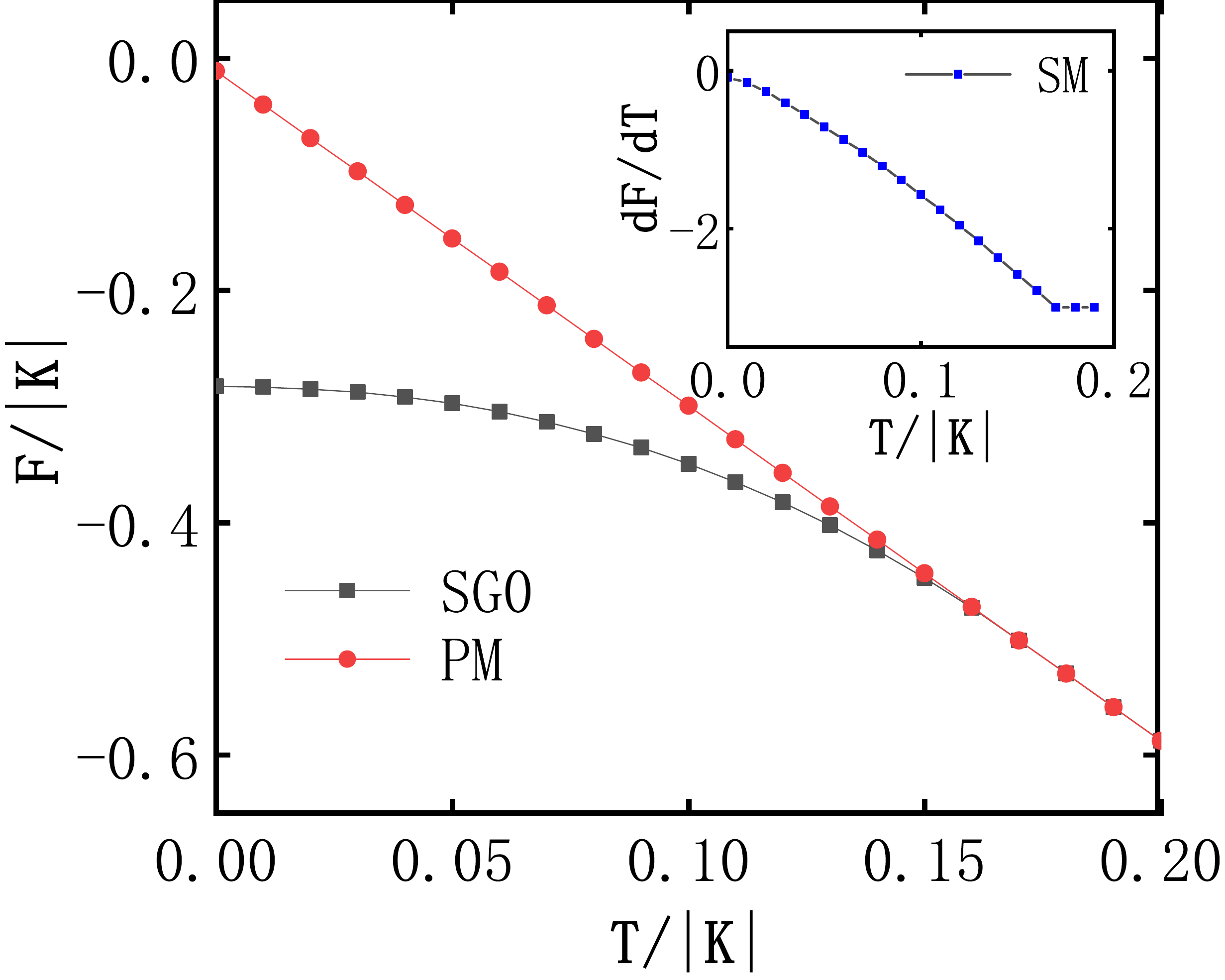}}
\caption{ Phase transitions of doped QSL14 (with $\Gamma/|K|=0.4, \Gamma^{'}/|K|=-0.02$) at finite temperature illustrated by the free energy curves.  (a) $\delta=0.1$, the level crossing indicates a first order transition from the SC0$_{\rm SL}$ to the SC0. (b) $\delta=0.02$, the converging of the two curves indicates a second order continuous phase transition from the SG0 phase to the PM phase. The insert illustrates the first order derivative of the free energy.}
\label{fig:E}
\end{figure}

\subsection{Finite Doping}

In this section, we study the physical consequence when holes are doped into the system.  Many interesting phases have been obtained, including (topological) superconductors, spin-gapped phases, the fermi liquid phase, the strange metal phase and the paramagnetic phase. Most of these phases appeared in the $t$-$J$ model as discussed in Ref.\cite{lee1992gauge, ubbens1992flux}, but our phase diagram is even richer [see Fig.\ref{figure2}(a)$\sim$(d)] owing to spin-orbit coupling in the spin-spin interactions.

{\bf Different mean field phases}.  In mean field theory, the fermionic spinons and the bosonic holons are decoupled, so the total free energy is a sum of that of the spinons and that of the holons. We tried several self-consist solutions, the one with the lowest free energy is adopt to describe the state of the system.

Here we provide the mean-field description of the phases listed above. The superconducting (SC) phase is characterized by the Bose-Einstein condensation of the holons and the condensation of the spinon pairs. If the spinon pairing parameter is nonzero but the holons are not condensed due to thermal fluctuations, then the resultant state belongs to the spin-gapped phase (SG) (also called the pseudogapped phase). On the other hand, if the holons condense but the spinon pairing parameter is zero, then the resultant phase is the fermi liquid (FL) phase. If holons are not condensed, and if the spinons do not form pairs but can coherently hop, then the resultant state is the strange metal (SM) phase. When both the spinons and the holons lose their coherence (namely, the all of the parameters $t_{0,1,2,3}$ and $\Delta_{0,1,2,3}$ vanish), and the system enters the high temperature paramagnetic (PM) phase.   
 
In the following we illustrate the steps of obtaining the phase diagram. 
 
Firstly we investigate the phase boundary of the magnetically ordered phases. In section \ref{sec:Dord}, we have estimated the critical temperature $T_{c\rm M}$ and the critical concentration $\delta_{c\rm M}$ (see Fig.\ref{fig:trans}(a) for example) for the magnetic transition. From $T_{c\rm M}$ and $\delta_{c\rm M}$, an  approximate phase boundary of the magnetic ordered phase can be figured out by drawing a smooth curve linking $T_{c\rm M}$ and $\delta_{c\rm M}$.

Then we estimate the critical temperature $T_{c\rm HC}$ for the holon condensation. In two-dimensions bosonic systems cannot spontaneously break continuous symmetries at any finite temperature.  In our study, the $T_{c\rm HC}$ stands for the Kosterlitz-Thouless transition temperature and is evaluated approximately\cite{kosterlitz1973ordering}. 

With the increasing of doping concentration $\delta$, another important feature is the critical concentration $\delta_{c\rm SP}$ at which the spinons pairing parameter drops to zero. To this end, we compare the ansatz (\ref{MFU}) with nonzero spinon-pairing (the SC states at lower $T$ or the SG states at higher $T$) and another ansatz (\ref{MFtD}) with vanishing spinon pairing terms (the FL state at lower $T$ or the SM state at higher $T$). It turns out that the latter ansatz is favored by the free energy when the doping concentration exceeds the critical $\delta_{c\rm SP}$. An example of such transition is shown in Fig.\ref{fig:trans}(a).

As mentioned in section \ref{sec:dopQSL}, different types of spinon ansatz have been considered to describe the SC phases. It turns out that at small $\delta$ (after the magnetic order being suppressed) and at low $T$, the spinon ansatz preserving the Kitaev PSG is lower in free energy. Above a critical $\delta$ or a critical $T$, the spinons prefer another pairing state outside the Kitaev PSG class. Fig.\ref{fig:E} (a) shows a first order transition between two SC phases with increasing $T$, where the low temperature SC0$_{\rm SL}$ preserves the Kitaev PSG and the high temperature SC0 does not.  Another example can be found in Fig.\ref{fig:trans}(a), where a first order transition between SC2$_{\rm zz}$ (preserving the Kitaev PSG) and SC0 (violating the Kitaev PSG) appears at a critical $\delta$ at zero temperature. Owing to this subtle difference, doped Kitaev system contains more than one SC phase and more than one pseudogap phase.

Finally, the paramagnetic phase is characterized by the vanishing of all parameters, in which the spinons and holons lose their coherence. There is a continuous phase transition between SM and PM (see Fig.\ref{fig:E} (b)).

From the above procedure, the global phase diagram is obtained. In the following, we analyze the doped ordered phases and doped QSLs one-by-one.  The doped KSL has been studied in Ref.~\cite{you2012doping} and will be skipped here.

{\bf Doped zigzag phase.}  The doped zigzag state is most interesting because it may be relevant to several Kitaev materials. Fig. \ref{figure2}(a) shows the phase diagram of doped zigzag phase with $\Gamma/|K|=1, \Gamma^{'}/|K|=-0.02$. The solid lines denote phase transitions obtained by 
minimizing the free energy of self-consistent mean-field solutions, 
 while the dotted lines are obtained by estimation using the method given previously.  

Interestingly, if the doping concentration is close to $\delta=0.1$,  we obtain a topological superconductor SC2$_{\rm zz}$ with Chern number $\nu = {2}$ or $\nu=-2$. In this region, the spinon states preserves the Kitaev PSG, and the nonzero Chern number results from the spontaneous breaking of time reversal symmetry in the complex parameter $\chi_b$ and $\chi_f$. The critical temperature of the superconductivity is around $0.08|K|$ (which is of the order of $10$ Kelvin for $\alpha$-RuCl$_3$\cite{JSWenPRL2017, HaiLiNC2021}). The SG2$_{\rm zz}$ phase appears when the temperature exceeds the $T_{c\rm HC}$, in which the holons are not condensed but the spinons stay in the same state as in the SC2$_{\rm zz}$ phase. With increasing $\delta\sim 0.2$, another superconductor SC0 with Chern number $\nu=0$ and the corresponding SG0 are obtained, in which the spinons still form pairs but are no longer preserving the Kitaev PSG. As $\delta>0.32$, the system becomes metalic and enters the FL or SM phase.

When the doping concentration reaches some value, the resultant physics is not sensitive to the spin-spin interactions. Therefore, the phase diagrams Fig.\ref{figure2}(a)$\sim$(d) in the region $\delta>0.14$ are very similar.

{\bf Doped QSL14.} Fig. \ref{figure2}(b) shows the phase diagram of doped QSL14 phase with $\Gamma/|K|=0.4, \Gamma^{'}/|K|=-0.02$. Since in the QSL14 phase the spinons have already formed pairs,  the system becomes superconducting once holes are doped with $0<\delta \leq 0.32$. 

Three different SC phases are obtained, one has Chern number $\nu=\pm{1}$ and the other two have $\nu=0$. The first two SC phases, SC1$_{\rm SL}$ and SC0$_{\rm SL}$ respect the Kiatev PSG and are separated at $\delta=0.06$ with a first-order phase transition (see Fig.\ref{fig:trans} (b)). In the third SC phase SC0, the spinon state falls outside the Kitaev PSG class and the pairing gap decreases with increasing $\delta$.

{\bf Doped FM and IS phases.} The phase diagrams of the doped FM  and IS states are shown in Fig.  \ref{figure2}(c) and (d), respectively, which are very similar to the case with doped zigzag state shown in Fig.\ref{figure2}(a). The main difference is that the Chern number of the topological SC phase doped from the FM and IS phases are $\nu=\pm{1}$.

\subsection{Effect of In-plane Magnetic Fields}

Below the magnetic transition temperature, the zigzag order in Kitaev materials (such as $\alpha$-RuCl$_3$) can be easily suppressed by in-plane magnetic field, resulting in a QSL-like disordered phase. In this section, we study the effect of in-plane magnetic field in the doped zigzag phase, which may be relevant for future experimental study. 

We consider the in-plane magnetic field, say, along the ${1\over\sqrt2} (\hat {\pmb x}-\hat {\pmb y})$ direction. In this case, the spinons gain Zeeman energy 
$$H^{\rm zeeman}_{\rm mf} =  {\mu_B B\over2\sqrt2}\sum_i  f_i^\dag (\sigma^x-\sigma^y)f_i,$$ 
but the cyclotron motion of the charge carriers is avoided. Thus the holons are not affected by the in-plane magnetic field.  Here we only discuss the situations at zero temperature. 

Since the in-plane magnetic field frustrates the zigzag order, it will need less doped holes to destroyed the order and to drive the system to the superconducting phase. Namely, the in-plane magnetic field can reduce the critical doping concentration $\delta_{c\rm M}$ to the SC phase. Similarly, doped holes also decreases the lower critical magnetic field strength in suppressing the magnetic order.

\begin{figure}[t]
\includegraphics[width=7cm]{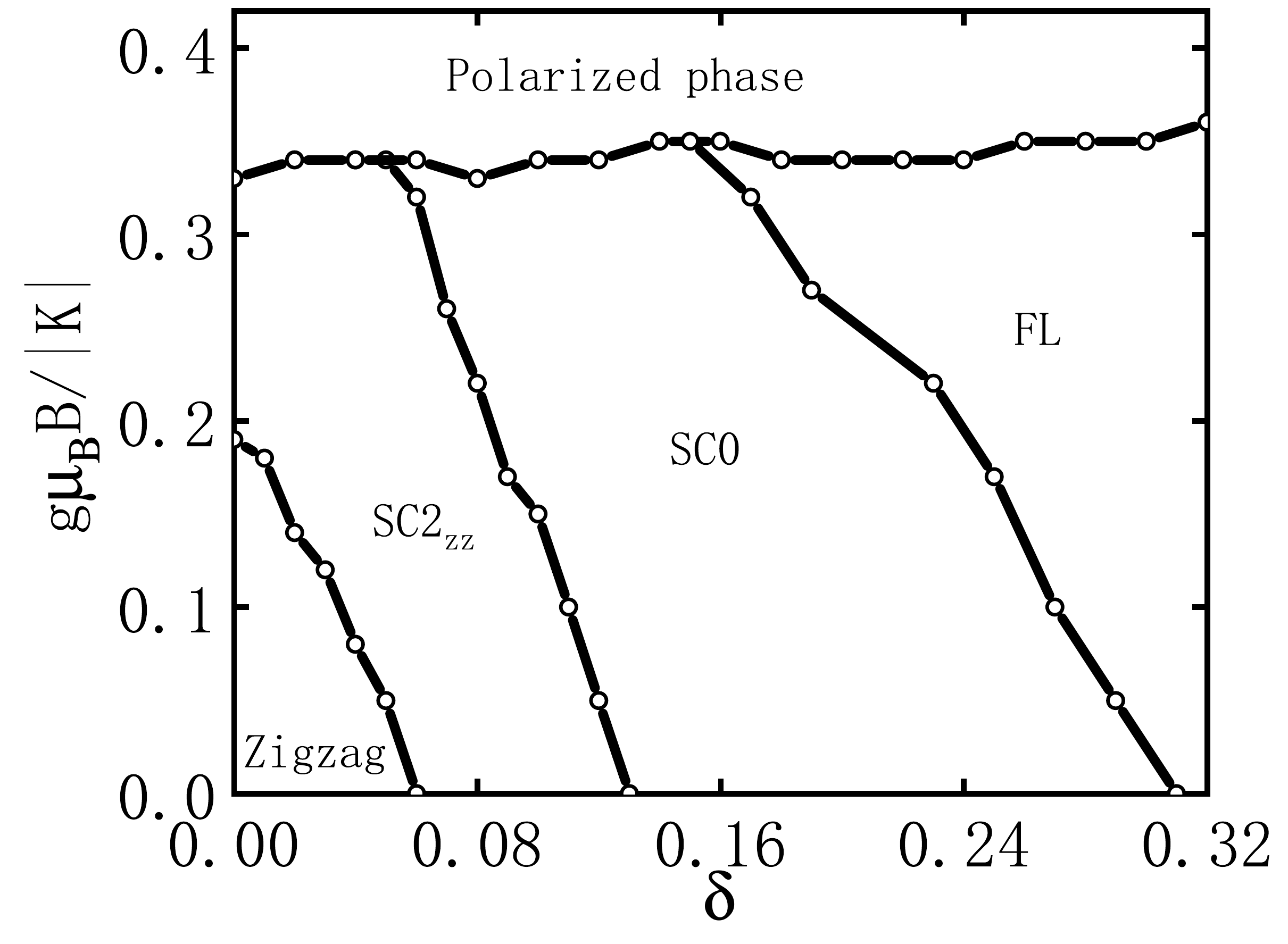}
\caption{Zero temperature phase diagram with in-plane magnetic fields $\pmb B\parallel {1\over\sqrt2}(\hat x-\hat y)$ applied to the hole-doped zigzag state. }\label{fig:B}
\end{figure}

On the other hand, the magnetic field tends to polarize the spinons and breaks the spinon pairing, so if the magnetic field reaches the upper critical value, the spin of the electrons will be polarized and the system enters the trivial polarized phase. For the same reason, the in-plane magnetic field results in a smaller critical $\delta_{c\rm SP}$ for the system to enter the fermi liquid phase. The phase diagram is shown in Fig.\ref{fig:B}. It can be seen that the size of the topological SC2$_{\rm zz}$ phase is slightly enlarged while the nontopological SC0 phase is shortened at the intermediate region of intensity of the magnetic field. 

It should be noted that vertical line with $\delta=0$ is special, where the intermediate phase between the Zigzag phase and the polarized phase is a gapless quantum spin liquid phase with two Majorana cones (similar results are obtained previously\cite{wang2020multinode} using variational Monte Carlo method). Once holes are doped into the system such that $\delta\neq 0$, the holons will condense and at the mean while the cones will open a gap, resulting in the topological superconducting phase SC2$_{\rm zz}$.

Comparing Fig.\ref{fig:B} and Fig.\ref{figure2}(a), it can be seen that the effect of in-plane magnetic field is very similar to that of the temperature. Similar results can be obtain in the doped FM phase or the doped IS phase and will not be shown here.

\section{Conclusions and Discussions}

In summary, to explore the effect of doped holes in Kitaev systems, we studied the $t$-$K$-$\Gamma$-$\Gamma^{'}$ model via mean field theory. Firstly we (almost) reproduce the zero temperature phase diagram of the pure spin model with $K$-$\Gamma$-$\Gamma^{'}$ interactions at zero doping, which contains three magnetically ordered phases and two QSL phases (the KSL and the QSL14 which contains 14 majorana cones).  By doping holes to these phases, we obtain the superconducting phases, the pseudogap phases, the fermi liquid phase, the strange metal phase and the paramagnetic phase. Topological superconductors are obtained, where the Chern number is dependent on the original spin state. By doping the zigzag  phase one has Chern number $\nu=\pm{2}$, by doping the FM phase or the IS phase one get $\nu=\pm{1}$. Interestingly, no matter what is the original spin state, more than one superconductor and more than one spin gapped state are obtained after doping. We studied the effect of applied in-plane magnetic field and find that it can slightly enlarge the size of the topological superconducting phase.

It should be clarified that the conclusions obtained from our mean field theory is very preliminary. Except for the magnetically ordered phases, we only consider the ansatz which preserving the $C_3$ and translation symmetry. Other symmetry breaking orders, such as the charge/spin density wave, nematicity, and valence bond solid order may appear at low temperatures but are not considered. Furthermore, quantum fluctuations around the mean field solution may quantitatively or qualitatively change the results. For instance, in the variational Monte Carlo phase diagram using Gutzwiller projected states as trial states,  a 14-cone QSL called PKSL14  (proximate Kitaev spin liquid with 14 cones) is obtained. But this PKSL14 is different from the QSL14 obtained from the self-consistent mean field solution, since the two states have different Chern number in a small magnetic field applied along $\pmb B\parallel {1\over\sqrt3}(\hat x+\hat y+\hat z)$ direction. Therefore, it will be interesting to study the present model using partially projected wave functions as trial states and determine the parameters variationally. We leave this for future study.

Finally, the $t$-$K$-$\Gamma$-$\Gamma'$ model is based on the single-band approximation. It is uncertain if this approximation is good to describes the low energy physics of doped Kitaev materials. If not, it deserves further study to find the true effective model at finite doping. 

Owing to the obtained rich phase diagram, our work may intrigue interest of future experimental studies.

\textit{Acknowledgement} -- We thank J.-C. Wang, Q.R. Zhao and F. Yang for valuable discussions. This work is supported by the Ministry of Science and Technology of China (Grant No. 2016YFA0300504), the NSF of China (Grants No. 11574392 and No. 11974421), and the Fundamental Research Funds for the Central Universities and the Research Funds of Renmin University of China (No. 19XNLG11).

\appendix
\setcounter{equation}{0}
\section{General form of Mean field Hamiltonian and projective symmetry groups}\label{app:dec}

In the main text, we introduce the fermionic spinon representation where the spin operators can be written as the familiar quadratic form of fermions $S^\alpha_i=\frac{1}{2}f^\dagger_i\sigma_\alpha f_i$ under the constraint $f^\dagger_{i } f_{i } = 1$, where $f^\dagger_i=(f^\dagger_{i\uparrow} , f^\dagger_{i\downarrow})$ and 
$\sigma_\alpha, \alpha=x,y,z$ are the Pauli matrix. The two spinon species may further be expressed in terms of four Majorana fermions,
\begin{align}
f_\uparrow=\frac{1}{2}(b^z+ic),\qquad f_\downarrow=\frac{1}{2}(b^x+ib^y),\nonumber
\end{align}
which satisfy the anti-commutation relations $\{b^\alpha,b^\beta\}=2\delta^{\alpha \beta}$ 
($\alpha,\beta =0,x,y,z;b^0\equiv c$). In this basis, the spin operator takes the form $S^m=ib^mc$.

There is a SU(2) gauge symmetry\cite{cookmeyer2018spin} in the spinon representation. To see this, we introduce $\bar{f}_i=(f^\dagger_{i\downarrow} -f^\dagger_{i\uparrow})^T$, the time reversal partner of $f^\dagger_i$, and a matrix operator $\psi_i=(f_i, \bar f_i)$. 

The mixing between $f$ and $\bar{f}$, namely, multiplying $\psi$ by a SU(2) matrix from the right hand side, does not change the spin operators $S^m_i={\rm Tr} (\psi^\dagger_i \frac{\sigma_m}{4} \psi_i)$, which defines the SU(2) gauge symmetry.

For rotationally invariant interactions. The Heisenberg exchange interaction can be written as
\beq
\pmb{S}_i\cdot \pmb S_j \!=\! -\frac{1}{8}{\rm Tr} (\psi^\dagger_j \psi_i \psi^\dagger_i \psi_j) 
\! =\! \frac{1}{8}{\rm Tr} (\psi^\dagger_j \bm{\sigma} \psi_i \cdot \psi^\dagger_i \bm{\sigma} \psi_j)
\label{heisenberg}
\eeq
up to constant terms. The anisotropic interactions, such as the AFM Ising interaction, can be decoupled as
\begin{align}
S^m_i S^m_j=-\frac{1}{16}[{\rm Tr} (\psi^\dagger_j \psi_i \psi^\dagger_i \psi_j)
                         +{\rm Tr} (\psi^\dagger_j \sigma^m \psi_i \psi^\dagger_i \sigma^m \psi_j)]
\end{align}
up to a constant term, where $m = x, y, z$.

To make more transparent the connection to the spinon representation, we note that the singlet matrix operator $\psi^\dagger_i \psi_j$, and the triplet operator $\psi^\dagger_i \bm{\sigma} \psi_j$,  can be expanded as
\begin{align}
\psi^\dagger_i \psi_j=
\begin{pmatrix} 
f^\dagger_i f_j & f^\dagger_i\bar{f}_j\\
\bar{f}^\dagger_if_j & \bar{f}^\dagger_i\bar{f}_j
\end{pmatrix},
\psi^\dagger_i \bm{\sigma} \psi_j=
\begin{pmatrix} 
f^\dagger_i \bm{\sigma} f_j & f^\dagger_i \bm{\sigma} \bar{f}_j\\
\bar{f}^\dagger_i \bm{\sigma} f_j & \bar{f}^\dagger_i \bm{\sigma} \bar{f}_j
\end{pmatrix},\nonumber
\end{align}
respectively.
\par

Actually, from above expression of the spin-spin interactions, we obtain the most general mean-field Hamiltonian (with only nearest neighbor couplings) for a spin-liquid state with spin-orbit coupling,
\beq
H^{\rm SL}_{\rm mf}&=&\sum_{ij} {\rm Tr} \Big(U^{(0)}_{ji}\psi^\dagger_i \psi_j + \sum_{m=x,y,z} {U}^{(m)}_{ji} \psi^\dagger_i  {\sigma^m} \psi_j + {\rm H.c.}\Big) \notag \\
&&+ \sum_i \pmb{\lambda} \cdot \pmb{\varLambda_i},
\label{HSL}
\eeq
where $U_{ji}^{(0)}$ is the same as that in (\ref{MFU}), and the matrices $U_{ji}^{(x,y,z)}$ are linearly related to the ones $U_{ji}^{(1,2,3)}$. For example, in the $z$-bond (namely for $\langle ij \rangle \in z$), one has
\Beq
U_{ji}^{(x)}&=&\frac{1}{\sqrt{2}}(U^{(1)}_{ji}-U^{(3)}_{ji}),
\\ U_{ji}^{(y)}&=&\frac{1}{\sqrt{2}}(U^{(1)}_{ji}+U^{(3)}_{ji}), 
\\ U_{ji}^{(z)}&=&U^{(2)}_{ji}.
\Eeq
When expanding the matrices $U_{ji}^{(0,1,2,3)}$ by the Pauli bases, the coefficients correspond to the parameters $t_{0,1,2,3}$ and $\Delta_{0,1,2,3}$ in (\ref{MFtD}). The advantage of the nation $U_{ji}^{(0,1,2,3)}$ is that if the mean field Hamiltonian preserves $C_3$ symmetry, then the values of $U_{ji}^{(0,1,2,3)}$ are independent on the bond directions.

The operators $\pmb \Lambda_i=\frac{1}{4}{\rm  Tr}(\psi_i \bm{\tau} \psi^\dagger_i)$ are the generators of the SU(2) gauge group, and the third component of $\pmb \lambda$ is the Lagrangian multiplier of the particle number constraint appeared in the main text. 

As shown in appendix \ref{app:psg}, above Hamiltonian (\ref{HSL}) is the general expression of the mean-field Hamiltonian even if the system contain off-diagonal interactions (with only nearest neighbor interactions).

The mean-field Hamiltonian of Eq. (\ref{HSL}) is not in general invariant under an arbitrary SU(2) gauge transformation. 
The subgroup of the SU(2) gauge group under which Eq. (\ref{HSL}) remains invariant is called the invariant gauge group (IGG) of the spin-liquid state. On the other hand, a QSL should respect all the symmetries of the spin Hamiltonian, but at the mean-field level this constraint can be relaxed in the following sense. Under a symmetry operation $g$, $H_{mf}$ may be transformed to a different expression, $\hat gH_{mf} \hat g^{-1}=H^{'}_{mf}\ne H_{mf}$, but if it can be transformed back to its original form by an SU(2) gauge transformation then this mean-field Hamiltonian still describes a spin-liquid state.  Specifically,
\begin{subequations}
\begin{align}
&\psi_i \to \hat g^\dagger \psi_{g(i)} W_i(g)\label{gW},
\\{\rm Tr} [U_{ji} \psi^\dagger_i \psi_j] &\to {\rm Tr} [W_j U_{ji} W^\dagger_i \psi^\dagger_{g(i)} \hat g\hat g^\dagger \psi_{g(j)}], \nonumber
\\&={\rm Tr} [U_{g(j)g(i)} \psi^\dagger_{g(i)} \psi_{g(j)}],
\\{\rm Tr} [\bm{U}_{ji} \cdot \psi^\dagger_i \bm{\sigma} \psi_j] &\to {\rm Tr} [W_j \bm{U}_{ji} W^\dagger_i \cdot \psi^\dagger_{g(i)} \hat g\bm{\sigma}\hat g^\dagger \psi_{g(j)}], \nonumber
\\&={\rm Tr} [\bm{U}_{g(j)g(i)} \cdot \psi^\dagger_{g(i)} \bm{\sigma} \psi_{g(j)}]
\end{align}
\end{subequations}
where $\hat g$ is the double-valued representation of $g$, $U^n_{g(j)g(i)}=\sum_m R_{nm}(g) W_j(g) U^m_{ji} W^\dagger_i(g)$, $U_{g(j)g(i)}=W_j(g) U_{ji} W^\dagger_i(g)$ and $R(g)$ is a vector representation of $g$. The new symmetry operations, each of which involves a symmetry operation $g$ followed by a gauge transformation $W_i(g)$, form a larger group that is known as the projective symmetry group (PSG)\cite{wang2019one,wang2020multinode}.

Now we provide the PSG of the Kitaev Spin Liquid (namely, the Kitaev PSG), which can be read out in the Majorana representation. The symmetry group $G = D_{3d}\times Z^T_2$ has three generators$$S_6=(C_3)^2P,\quad M=C^{x-y}_2 P,\quad T=i\sigma^yK,$$
where $C_3$ is a threefold rotation around the direction $\hat{c}\equiv \frac{1}{\sqrt{3}} (\hat{x}+\hat{y}+\hat{z})$, $C^{x-y}_2$ is a twofold rotation around $\frac{1}{\sqrt{2}}(\hat{x}-\hat{y})$, and $P$ is spatial inversion. From (\ref{gW}), one obtain the Kitaev PSG, 
\Beq
&&W_A(S_6) = -W_B(S_6) = \exp[-i\frac{4\pi}{3} \frac{1}{2\sqrt{3}}(\tau^x+\tau^y+\tau^z)],\nonumber
\\&&W_A(M) = -W_B(M) = \exp[-i\pi\frac{1}{2\sqrt{2}}(\tau^x-\tau^y)],\nonumber
\\&&W_A(T) = -W_B(T)=i\tau^y,\nonumber
\Eeq
where A and B denote the two sublattices of the honeycomb lattice.

\section{Decoupling of the $K$-$\Gamma$-$\Gamma'$ interactions}\label{app:psg}
In the Majorana representation, the mean-field decouplings of the Kitaev interactions which preserve the Kitaev PSG read
\begin{align}
H^K_{mf}&=\sum_{\langle i,j \rangle \in \alpha \beta (\gamma)} \rho_a(ic_ic_j)+\rho_c(ib^\gamma_ib^\gamma_j) \nonumber
\\&=\sum_{\langle i,j \rangle \in \alpha \beta (\gamma)} i\rho_a{\rm Tr} (\psi^\dagger_i \psi_j + \tau^x \psi^\dagger \sigma^x \psi_j 
+ \tau^y \psi^\dagger_i \sigma^y \psi_j \nonumber
\\& \qquad \quad \quad+\tau^z \psi^\dagger \sigma^z \psi_j) 
+i\rho_c{\rm Tr} (\psi^\dagger_i \psi_j + \tau^\gamma \psi^\dagger \sigma^\gamma \psi_j \nonumber
\\& \qquad \quad \quad- \tau^\alpha \psi^\dagger_i \sigma^\alpha \psi_j - \tau^\beta \psi^\dagger \sigma^\beta \psi_j) + {\rm H.c.}
\label{HK}
\end{align}
\par

Similarly, the $\Gamma$ interaction decouples as 
\begin{align}
H^\Gamma_{mf}&=\sum_{\langle i,j \rangle \in \alpha \beta (\gamma)} i\rho_d(b^\alpha_i b^\beta_j + b^\beta_i b^\alpha_j) \nonumber
\\&=\sum_{\langle i,j \rangle \in \alpha \beta (\gamma)} i\rho_d {\rm Tr} (\tau^\alpha \psi^\dagger_i \sigma^\beta \psi_j 
+ \tau^\beta \psi^\dagger_i \sigma^\alpha \psi_j) + {\rm H.c.}
\label{HT}
\end{align}
and for the $\Gamma^{'}$ interaction
\begin{align}
H^{\Gamma^{'}}_{mf}&=\sum_{\langle i,j \rangle \in \alpha \beta (\gamma)} i\rho_f (b^\alpha_i b^\gamma_j + b^\gamma_i b^\alpha_j
+ b^\beta_i b^\gamma_j + b^\gamma_i b^\beta_j) \nonumber
\\&=\sum_{\langle i,j \rangle \in \alpha \beta (\gamma)} i\rho_f {\rm Tr} (\tau^\alpha \psi^\dagger_i \sigma^\gamma \psi_j 
+ \tau^\gamma \psi^\dagger_i \sigma^\alpha \psi_j \nonumber
\\& \qquad \qquad+ \tau^\beta \psi^\dagger_i \sigma^\gamma \psi_j + \tau^\gamma \psi^\dagger_i \sigma^\beta \psi_j) + {\rm H.c.}
\label{HTT}
\end{align}
However, the most general coefficients preserving the $C_3$ rotation symmetry also contain multiples of the uniform ($\tau^0$)
and $\tau^x + \tau^y + \tau^z$ gauge components. Thus we need to add extra terms in the mean-field Hamiltonian as long as they are symmetry allowed. The terms we found that preserve all the Kitaev PSG symmetries contain three more parameters $\eta_{0,1,2}$,
\begin{align}
H^{\rm ex}_{\rm mf}\!&=\!\!\!\!\sum_{\langle i,j \rangle \in \alpha \beta (\gamma)}\!\!\! i\eta_0 {\rm Tr}(\psi^\dagger_i \psi_j)\! +\! i\eta_2 {\rm Tr}[(\tau^x \!+\!\tau^y \!+\! \tau^z)\psi^\dagger_i\sigma^\gamma\psi_j]\nonumber
\\& \qquad \qquad + i\eta_1 {\rm Tr}[(\tau^x+\tau^y+\tau^z)\psi^\dagger_i(iR^\gamma_{\alpha\beta})\psi_j]+{\rm H.c.}
\label{MF}
\end{align}
Combining the equations (\ref{HK}), (\ref{HT}), (\ref{HTT}) and (\ref{MF}), we obtain the complete mean-field Hamiltonian (\ref{MFU}) and (\ref{prmrel}) which describe QSL states preserving the Kitaev PSG.

\bibliography{references}

\end{document}